\documentclass[11pt]{article}
\usepackage{times}
\usepackage{mathfont}
\usepackage{graphicx}
\usepackage{cite}
\usepackage{url}
\def\max{\mathop{\rm max}}

\mathcode`O="724F

\DeclareSymbolFont{lasy}{U}{lasy}{m}{n}
\let\Box\undefined
\DeclareMathSymbol\Box{0}{lasy}{"32}

\newtheorem{lemma}{Lemma}
\newtheorem{corollary}{Corollary}
\newtheorem{theorem}{Theorem}

\newcommand{\qed}{~$\Box$\medbreak}
\newenvironment{proof}{\noindent{\bf Proof: }}{\qed}

\begin{document}

\title{Improved Algorithms for $3$-Coloring,\\ $3$-Edge-Coloring, and
Constraint Satisfaction}

\author{David Eppstein\thanks{Dept. Inf. \& Comp. Sci., Univ. of
California, Irvine, CA 92697-3425.  Email: {\tt eppstein@ics.uci.edu}.}}

\date{ }
\maketitle

\begin{abstract}
We consider worst case time bounds for NP-complete problems including
$3$-SAT, $3$-coloring, $3$-edge-coloring, and $3$-list-coloring.  Our
algorithms are based on a constraint satisfaction (CSP) formulation of
these problems; $3$-SAT is equivalent to $(2,3)$-CSP
while the other problems above are special cases of $(3,2)$-CSP.  We
give a fast algorithm for $(3,2)$-CSP and use it to improve the
time bounds for solving the other problems listed above.
Our techniques involve a mixture of Davis-Putnam-style backtracking with
more sophisticated matching and network flow based ideas.
\end{abstract} 

\section{Introduction}

There has recently been growing interest in analysis of
superpolynomial-time algorithms, including algorithms for NP-hard problems
such as satisfiability or graph coloring.  
This interest has multiple causes:

\begin{itemize}
\item Many important applications can be modeled with these problems,
and with the increased speed of modern computers, solved effectively;
for instance it is now routine to solve hard 500-variable satisfiability
instances, and structured instances with up to 10000 variables can often
be handled in practice~\cite{Sel-FXS-00}.

\item Improvements in exponential time bounds are especially critical in
determining the size of problems that can be solved: an improvement
from $O(2^{c_1 n})$ to $O(2^{c_2 n})$ implies a factor of $c_1/c_2$
improvement in the solvable problem size while technological
developments can only improve the size by an additive constant.

\item Approximation algorithms for many of these problems are often
either nonsensical (how does one approximate SAT?) or inadequate.

\item The large gap between the known theoretical worst case bounds and
results from empirical testing of implementations
provides an interesting challenge to algorithm researchers.
\end{itemize}

In this paper we continue our previous work on exact
algorithms for $3$-coloring, $3$-edge-coloring, and
$3$-SAT~\cite{BeiEpp-FOCS-95}. Each of these problems can be expressed as a
form of {\em constraint satisfaction} (CSP).  We solve instances of CSP
with at most two variables per constraint by showing that such an
instance either contains a {\em good local configuration} allowing us to
split the problem into several smaller instances, or can be solved
directly by graph matching.  We solve graph $3$-coloring by using
techniques including network flow to find a small set of vertices with
many neighbors, choosing colors for that set of vertices, and treating
the remaining problem using our constraint satisfaction algorithm.  We
solve graph $3$-edge-coloring by a further level of case analysis: we use
graph matching to find a large set of good local configurations, each of
which can be applied independently yielding a set of instances of a
generalized edge coloring problems in which certain pairs of edges are
constrained to have distinct colors.  We then solve this generalized
coloring problem with our vertex coloring algorithm.

Beyond simply adding additional case analysis, the improvements in the
present work stem from the following new ideas:

\begin{itemize}
\item Extending our previous $(3,2)$-CSP algorithm to $(4,2)$-CSP,
and measuring the size of a $(4,2)$-CSP instance in terms of a
parameter $\epsilon$ which can be varied to achieve the optimal tradeoff
between different cases in the analysis.
\item Stopping the search when a CSP instance can be solved by a graph
matching algorithm, rather than continuing the case analysis of an
instance until it can be determined directly to be solvable or unsolvable.
\item Eliminating cycles of low-degree vertices from vertex coloring
instances, in order to show that a large fraction of the graph can be
covered by the neighborhoods of few high degree vertices.
\item Using network flow techniques to cover a vertex coloring instance
with a forest that avoids certain bad kinds of trees. Our previous paper
instead performed a similar step using a complicated case analysis in
place of network flow, and achieved weaker limitations on the types of
trees occurring in the forest.
\item Introducing a generalization of edge coloring,
so that we can perform reductions while staying in the same problem
class before treating the problem as an instance of vertex coloring,
and using graph matching to find many
independent good local configurations in an edge coloring instance.
\end{itemize}

We omit most of the case analysis and proofs in this extended
abstract.  For details see the full paper \cite{cs.DS/0006046},
which combines these new results with those from our previous conference
paper~\cite{BeiEpp-FOCS-95}.

\subsection{New Results}

We show the following:

\begin{itemize}
\item A $(3,2)$-CSP instance with $n$ variables can be solved in worst
case time $O(1.3645^n)$, independent of the number of constraints.
We also give a very simple randomized algorithm for solving this
problem in expected time $O(n^{O(1)} 2^{n/2})\approx O(1.4142^n)$.

\item A $(d,2)$-CSP instance with $n$ variables and $d>3$ can be solved by
a randomized algorithm in expected time $O((0.4518d)^n)$.

\item $3$-coloring in a graph of $n$ vertices can be solved in time
$O(1.3289^n)$, independent of the number of edges in the graph.

\item $3$-list-coloring (graph coloring given a list at each vertex of
three possible colors chosen from some larger set) can be solved in time
$O(1.3645^n)$, independent of the number of edges.

\item $3$-edge-coloring in an $n$-vertex graph can be solved in time
$O(2^{n/2})$, again independent of the number of edges.

\item $3$-satisfiability of a formula with $t$ $3$-clauses can be solved in
time $O(n^{O(1)}+1.3645^t)$, independent of the number of variables or
2-clauses in the formula.
\end{itemize}

Except where otherwise specified, $n$ denotes the number of
vertices in a graph or variables in a SAT or CSP instance, while
$m$ denotes the number of edges in a graph, constraints in an CSP
instance, or clauses in a SAT problem.

\subsection{Related Work}

For three-coloring, we know of several relevant references.
Lawler~\cite{Law-IPL-76} is primarily concerned with the general chromatic
number, but also gives a simple $O(1.4422^n)$-time algorithm for
$3$-coloring: for each maximal independent set, test whether the
complement is bipartite. Schiermeyer~\cite{Sch-WG-93} improves this to
$O(1.415^n)$, and our previous conference paper~\cite{BeiEpp-FOCS-95}
further reduced the time bound to $O(1.3443^n)$.
Our $O(1.3289^n)$ bound significantly improves
all of these results.

There has also been some related work on approximate or heuristic
$3$-coloring algorithms. Blum and Karger~\cite{BluKar-IPL-97} show that
any $3$-chromatic graph can be colored with $\tilde{O}(n^{3/14})$ colors in
polynomial time. Alon and Kahale~\cite{AloKah-SJC-97} describe a technique
for coloring random $3$-chromatic graphs in expected polynomial time, and
Petford and Welsh~\cite{PetWel-DM-89} present a randomized algorithm for
$3$-coloring graphs which also works well empirically on random graphs
although they prove no bounds on its running time. Finally,
Vlasie~\cite{Vla-TAI-95} has described a class of instances which are
(unlike random $3$-chromatic graphs) difficult to color.

Several authors have described exact algorithms for Boolean
formula
satisfiability~\cite{Dan-JSM-83,DanHir-TR-00,DavPut-JACM-60,
GraHirNie-SAT-00,Hir-SODA-98,Luc-FC-84,KulLuc-ms-94,Kul-TCS-99,
MonSpe-DAM-85,PatPudSak-FOCS-98,Rod-AISMC-96,Sch-CSL-92}.
Very recently, Sch\"oning~\cite{Sch-FOCS-99} has described a simple and
powerful randomized algorithm for $k$-SAT and more general constraint
satisfaction problems, including the CSP instances that we use in our
solution of $3$-coloring.
For $3$-SAT, Sch\"oning's algorithm takes expected time
$O((4/3+\epsilon)^n)$ However, for $(d,2)$-CSP, Sch\"oning notes that
his method is not as good as a randomized approach based on an idea from
our previous conference paper~\cite{BeiEpp-FOCS-95}: simply
choose a random pair of values for each variable and solve the resulting
2-SAT instance in polynomial time, giving an overall bound of
$O((d/2)^n n^{O(1)})$.
Feder and Motwani~\cite{FedMot-98} have an alternative randomized
algorithm which takes time $O((d!)^{n/d} n^{O(1)})$, an improvement
over Sch\"oning for
$d\ge 6$ (and over our results for $d\ge 11$). The table below compares
these bounds with our new results; an entry with value $x$ in column $d$
indicates a time bound of
$O(x^n n^{O(1)})$ for $(d,2)$-CSP.

\begin{table}[h]
\begin{center}
\begin{tabular}{|l|l|l|l|l|l|l|} \hline
&$d=3$&$d=4$&$d=5$&$d=6$&$d=7$&$d=8$\\ \hline
Sch\"oning~\cite{Sch-FOCS-99}&1.5&2&2.5&3&3.5&4\\ \hline
Feder and Motwani~\cite{FedMot-98}&1.8171&2.2134&2.6052&2.9938%
&3.3800&3.7644\\ \hline
New results&1.3645&1.8072&2.2590&2.7108&3.1626&3.6144\\ \hline
\end{tabular}
\end{center}
\end{table}

The only prior work we found for $3$-edge coloring was our own
$O(1.5039^n)$ bound~\cite{BeiEpp-FOCS-95}. Since any $3$-edge-chromatic
graph has at most
$3n/2$ edges, one can transform the problem to $3$-vertex-coloring at the
expense of increasing
$n$ by a factor of $3/2$.  If we applied our vertex coloring algorithm we
would then get time $O(1.5319^n)$. Both of these bounds are significantly
improved by the one we state.

It is interesting that, historically, until the work of
Sch\"oning~\cite{Sch-FOCS-99}, the time bounds for $3$-coloring have been
smaller than those for $3$-satisfiability (in terms of the number of
vertices or variables respectively).  Sch\"oning's $O((4/3+\epsilon)^n)$
time bound for
$3$-SAT reversed this pattern by being smaller than the previous
$O(1.3443^n)$ bound for $3$-coloring from our
previous paper~\cite{BeiEpp-FOCS-95}.  The present work restores
$3$-coloring to a smaller time bound than $3$-SAT.

\section{Constraint Satisfaction Problems}

We now describe a common generalization of satisfiability and graph
coloring as a {\em constraint satisfaction problem}
(CSP)~\cite{Kum-AIM-92,Sch-FOCS-99}. We are given a collection of $n$
variables, each of which has a list of possible colors allowed.
We are also given a collection of $m$ {\em constraints},
consisting of a tuple of variables and a color for each variable.
A constraint is {\em satisfied} by a coloring if not every variable in
the tuple is colored in the way specified by the constraint. We would
like to choose one color from the allowed list of each variable, in a way
not conflicting with any constraints.

\begin{figure}[t]
$$\includegraphics[width=2in]{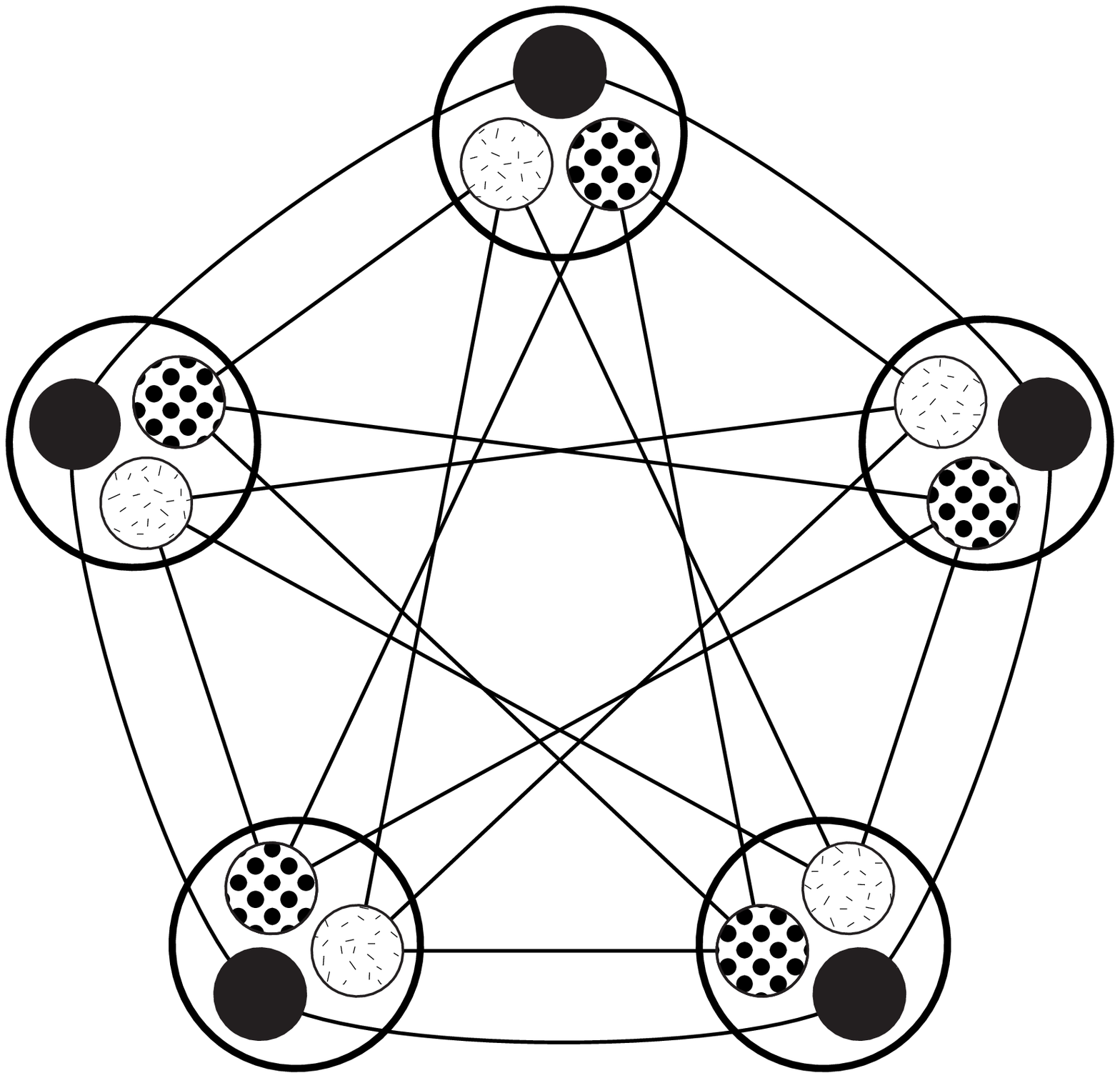}\qquad
\includegraphics[width=2in]{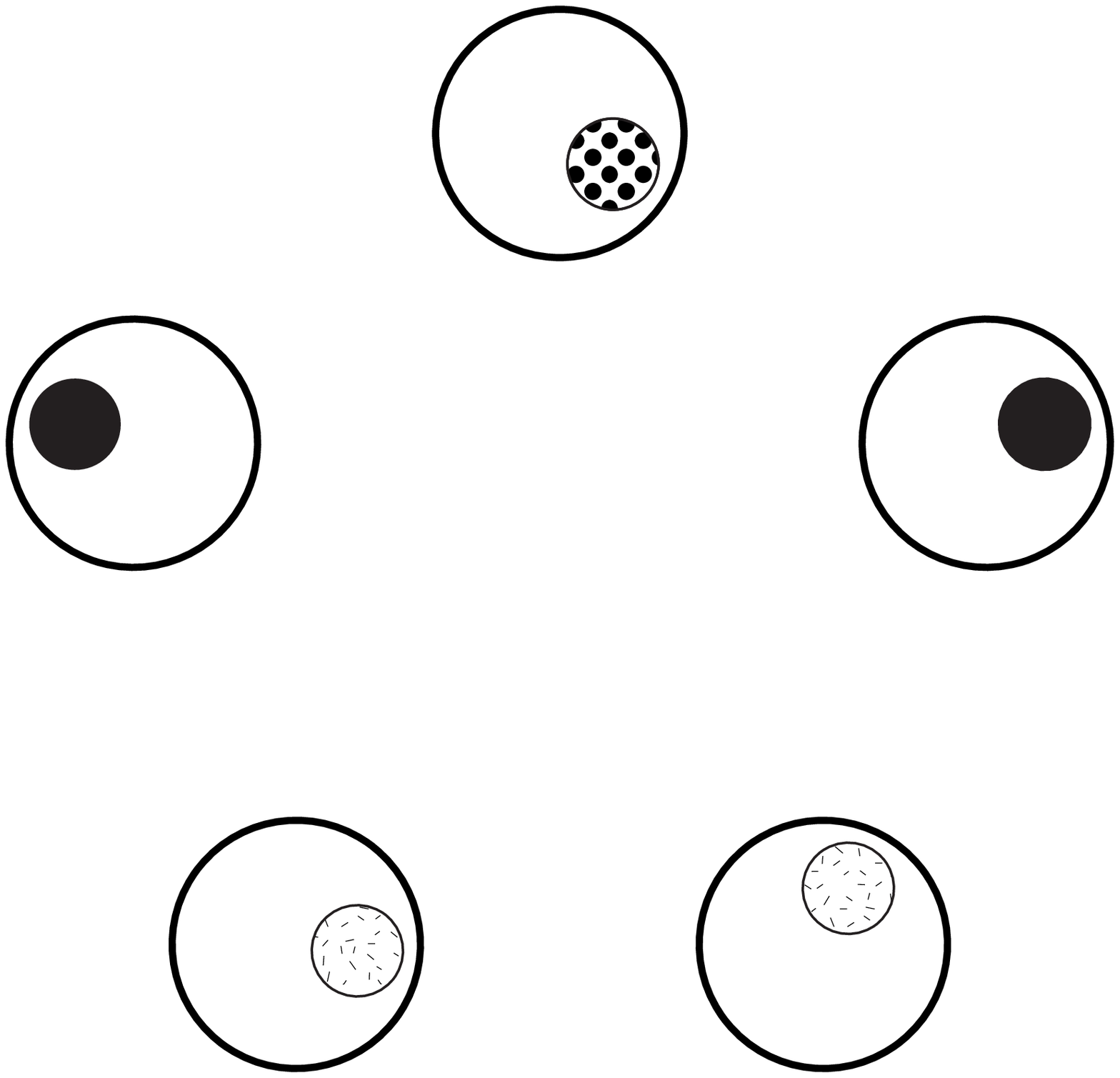}$$
\caption{Example $(3,2)$-CSP instance with five variables and twenty
constraints (left), and a solution of the instance (right).}
\label{fig:32sss}
\end{figure}

\begin{figure}[t]
$$\includegraphics[width=2.25in]{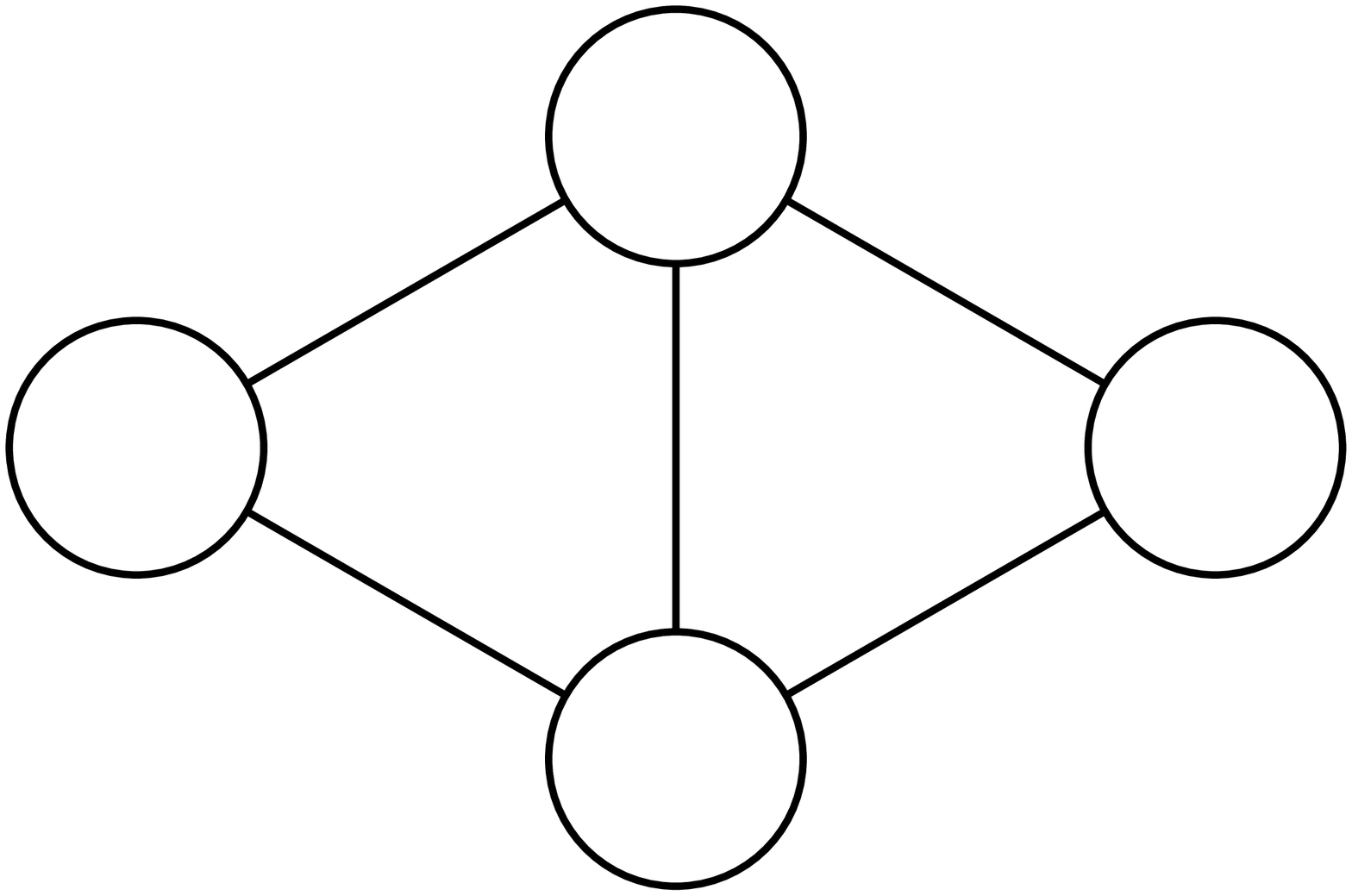}\qquad
\includegraphics[width=2.25in]{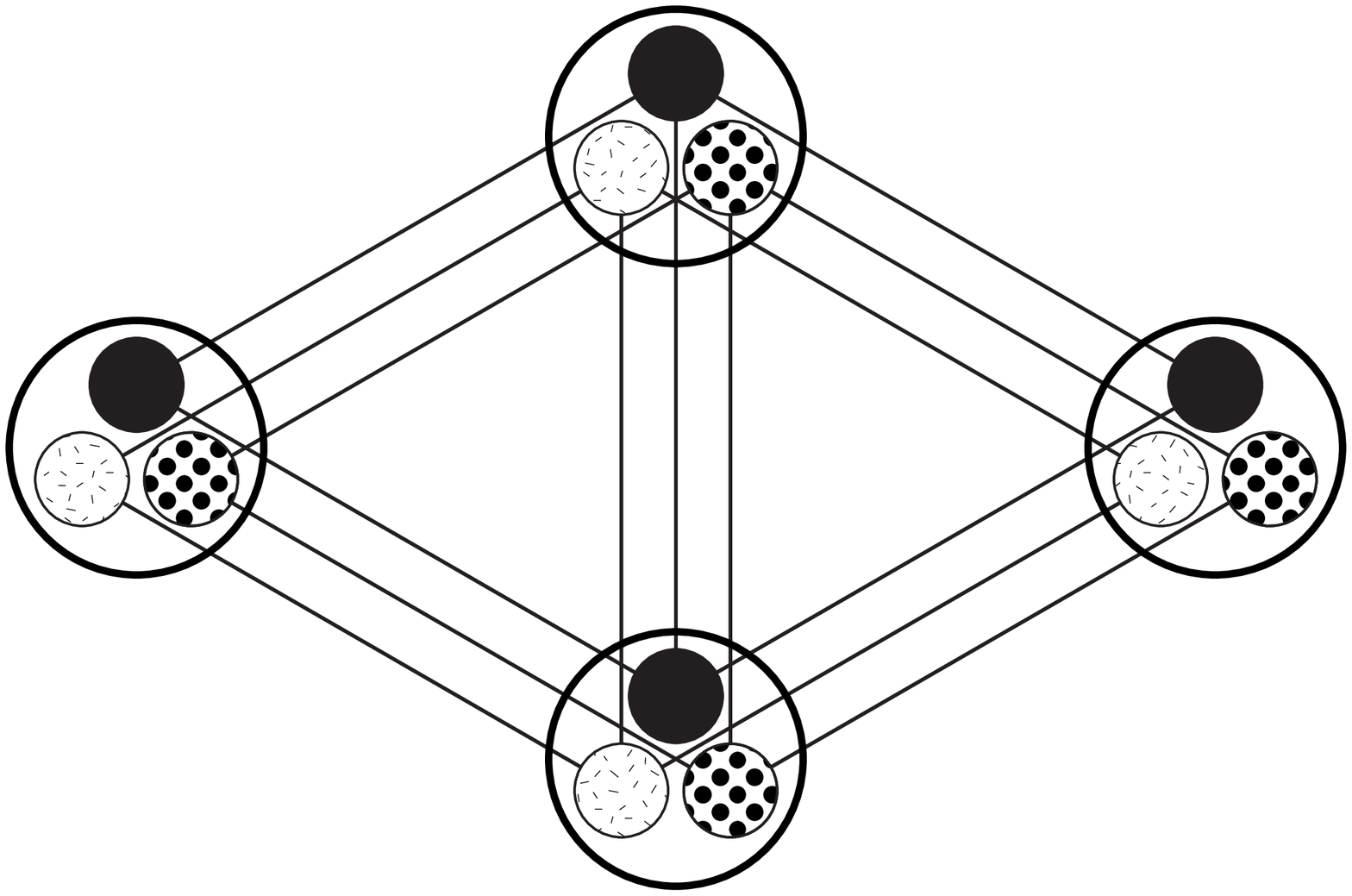}$$
\caption{Example $3$-coloring instance (left)
and translation into a $(3,2)$-CSP instance (right).}
\label{fig:gc2sss}
\end{figure}

For instance, $3$-satisfiability can easily be expressed in this form.
Each variable of the satisfiability problem may be colored (assigned the
value) either {\em true} ($T$) or {\em false}
($F$). For each clause like $(x_1 \vee x_2 \vee \neg{x}_3)$,
we make a constraint $((v_1,F),(v_2,F),(v_3,T))$.
Such a constraint is satisfied if and only if at least one of the
corresponding clause's terms is true.

In the {\em $(a,b)$-CSP} problem,
we restrict our attention to instances in which each variable has at most
$a$ possible colors and each constraint involves at most $b$ variables.
The CSP instance constructed above from a $3$-SAT instance is then
a $(2,3)$-CSP instance, and in fact $3$-SAT is easily seen to be
equivalent to $(2,3)$-CSP.

In this paper, we will concentrate our
attention instead on $(3,2)$-CSP and $(4,2)$-CSP.  We can represent a
$(d,2)$-CSP instance graphically, by interpreting each variable as a
vertex containing up to $d$ possible colors, and by drawing edges
connecting incompatible pairs of vertex colors (Figure~\ref{fig:32sss}). 
Note that this graphical structure is not actually a graph, as the edges
connect colors within a vertex rather than the vertices themselves.
However, graph $3$-colorability and graph $3$-list-colorability
can be translated directly to a form of $(3,2)$-CSP: we keep the
original vertices of the graph and their possible colors, and add up to
three constraints for each edge of the graph to enforce the condition
that the edge's endpoints have different colors
(Figure~\ref{fig:gc2sss}).

Of course, since these problems are all NP-complete, the theory of
NP-completeness provides translations from one problem to the other,
but the translations above are size-preserving and very simple. Our
graph coloring techniques include more complicated translations in which
the input graph is partially colored before treating the remaining graph
as an CSP instance, leading to improved time bounds over our pure CSP
algorithm.

\section{Constraint Satisfaction Algorithm}

\begin{figure}[t]
$$\includegraphics[width=3in]{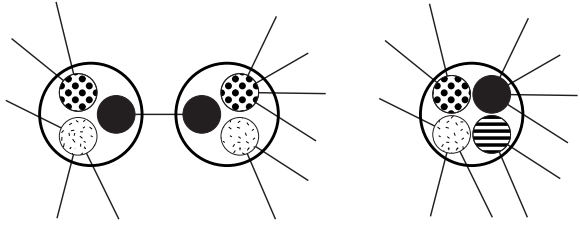}$$
\caption{Isolated constraint between two three-color variables (left) is
equivalent to a single four-color variable (right).}
\label{fig:iso33}
\end{figure}

We now outline our $(4,2)$-CSP algorithm.
A $(4,2)$-CSP instance can be transformed into a $(3,2)$-CSP instance
by expanding its four-color variables to two three-color
variables (Figure~\ref{fig:iso33}), so a natural definition of the
``size'' of a $(4,2)$-CSP instance is
$n=n_3+2n_4$, where $n_i$ denotes the number of $i$-color variables.
However, we instead define the size as
$n=n_3+(2-\epsilon)n_4$, where $\epsilon$ is a constant to be determined
later.  The size of a
$(3,2)$-CSP instance remains equal to its number of variables, so any
bound on the running time of our algorithm in terms of $n$ applies
directly to $(3,2)$-CSP.

\def\rec#1{\lambda(#1)}

Our basic idea is to find a set of local configurations
that must occur within any $(4,2)$-CSP instance $I$.
For each configuration we describe a
set of smaller instances $I_i$ of size $|I|-r_i$ such that $I$ is solvable
if and only if at least one of the instances $I_i$ is solvable.  If one
particular configuration occurred at each step of the algorithm, this
would lead to a recurrence of the form
$$T(n)=\sum T(n-r_i)+\hbox{poly}(n)=O(\rec{r_1,r_2,\ldots}^n)$$
for the runtime of our algorithm, where
$\rec{r_1,r_2,\ldots}$ is the largest
zero of the function
$f(x)=1-\sum x^{-r_i}$.
We call this value $\rec{r_1,r_2,\ldots}$
the {\em work factor} of the given local configuration.
The overall time bound will be $\lambda^n$ where $\lambda$ is the largest
work factor among the configurations we identify.

We first consider local configurations in which some (variable,color)
pair $(v,R)$ is involved in only a single constraint $((v,R),(w,R))$.
If this is also the only constraint involving $(w,R)$,
and both $v$ and $w$ have three colors, they
can be replaced by a single four-color variable
(Figure~\ref{fig:iso33}); any other singly-constrained color leads
to a problem reduction with work factor
$\rec{2-\epsilon,$3$-\epsilon}$.

We next find colors with multiple constraints to different colors
of the same variable, and show that such a case has work factor
$\rec{2-\epsilon,$3$-2\epsilon}$.
The next case we consider involves colors constrained by four or more
neighboring variables, or four-color variables with a color constrained
by three variables.  In these cases, choosing to use or not use
the highly-constrained color gives work factor
$\rec{1-\epsilon,5-4\epsilon}$.
Instances in which none of the above cases applies have a special form:
each (variable,color) pair has exactly two or three constraints, which
must involve distinct variables.

Our next sequence of cases concerns adjacency between (variable,color)
pairs with two constraints and pairs with three constraints.
We show that, if $(v,R)$ has three constraints,
one of which connects it to a variable with four color choices,
then the
instance can be replaced by smaller instances with work factor at most
$\rec{$3$-\epsilon,4-\epsilon,4-\epsilon}$.
If this case does not apply, and a three-constraint pair $(v,R)$
is adjacent to a (variable,color) pair with two constraints,
then we have additional cases with work factor at most
$\max\{\rec{1+\epsilon,4},\rec{3,4-\epsilon,4}\}$.
If none of these cases applies to an instance, then each color choice
in the instance must have either two or three constraints, and each
neighbor of that choice must have the same number of constraints.

We now consider
the remaining (variable,color) pairs that have three constraints each. 
Define a {\em three-component} to be a subset of such pairs such that any
pair in the subset is connected to any other by a path of constraints.  We
distinguish two such types of components: a {\em small three-component} is
one that involves only four distinct variables, while a {\em large
three-component} involves five or more variables.  
A small three-component is {\em good} if it involves only four
(variable,color) pairs.
We show that an instance containing a small three-component that is not
good can be replaced by smaller instances with work factor at most
$\rec{4,4,4}$, and that an instance containing a large three-component
can be replaced by smaller instances with work factor at most
$\rec{4,4,5,5}$.  As a consequence, we can assume all remaining
three-components are good.

Finally, we define a {\em two-component} to be a subset
of (variable,color) pairs such that each has two constraints, and any pair
in the subset is connected to any other by a path of constraints.
Our analysis of two-components is essentially the same as
in~\cite{BeiEpp-FOCS-95}, and shows that unless a two-component forms a
triangle, the instance can be replaced by smaller
instances with work factor at most $\rec{3,3,5}$.

Suppose we have a $(4,2)$-CSP instance to which none of the preceding
reduction cases applies.
Then, every constraint must be part of a good three-component or a
triangular two-component.  As we now show, this simple structure enables
us to solve the remaining problem quickly.

\begin{lemma}\label{lem:matching}
If we are given a $(4,2)$-CSP instance in which every constraint must be
part of a good three-component or a small two-component,
then we can solve it or determine that it is not solvable in polynomial
time.
\end{lemma}

\begin{proof}
We form a bipartite graph, in which the vertices correspond to the
variables and components of the instance.  We connect a variable to a
component by an edge if there is a (variable,color) pair using that
variable and belonging to that component. The instance is solvable iff
this graph has a matching covering all variables.
\end{proof}

This completes the analysis needed for our result.

\begin{theorem}\label{thm:sss}
We can solve any $(3,2)$-CSP instance in time 
$O(\rec{4,4,5,5}^n)\approx O(1.36443^n)$.
\end{theorem}

\begin{proof}
We employ a backtracking (depth first) search in a state space consisting
of $(3,2)$-CSP instances.  At each step we examine the
current state, match it to one of the cases above, and
recursively search each smaller instance.  If we reach an
instance in which Lemma~\ref{lem:matching} applies, we perform a matching
algorithm and either stop with a solution or backtrack to the most recent
branching point of the search and continue with the next alternative.

A bound of $\lambda^n$ on the number of recursive calls in this
search algorithm, where $\lambda$ is the maximum work factor occurring in
our reduction lemmas, can be proven by induction on the size of
an instance. To determine the maximum work factor, we need to set the
parameter $\epsilon$.  We used {\em Mathematica} to optimize $\epsilon$
numerically, and found that for $\epsilon\approx 0.095543$
the work factor is $\approx 1.36443\approx\rec{4,4,5,5}$.  For $\epsilon$
near this value, the largest work factors involving $\epsilon$ are
$\rec{$3$-\epsilon,4-\epsilon,4-\epsilon}$,  and
$\rec{1+\epsilon,4}$; the remaining work
factors are below 1.36.  The true optimum value of $\epsilon$ is thus the
one for which $\rec{$3$-\epsilon,4-\epsilon,4-\epsilon}=\rec{1+\epsilon,4}$.

As we now show, for this optimum $\epsilon$,
$\rec{$3$-\epsilon,4-\epsilon,4-\epsilon}=\rec{1+\epsilon,4}=\rec{4,4,5,5}$,
which also arises as a work factor in another case.
Consider subdividing an instance of size $n$ into one of
size $n-(1+\epsilon)$ and another of size $n-4$, and then further
subdividing the first instance into subinstances of size
$n-(1+\epsilon)-($3$-\epsilon)$, $n-(1+\epsilon)-(4-\epsilon)$, and
$n-(1+\epsilon)-(4-\epsilon)$. This four-way subdivision 
has work factor $\rec{4,4,5,5}$, and
combines
subdivisions of type $\rec{1+\epsilon,4}$ and
$\rec{$3$-\epsilon,4-\epsilon,4-\epsilon}$,
so these three work factors must be equal.
\end{proof}

We use the quantity $\rec{4,4,5,5}$ frequently in the remainder of the
paper, so we use $\Lambda$ to denote this value.
Theorem~\ref{thm:sss} immediately gives algorithms for some more well
known problems, some of which we improve later.
Of these, the least familiar is likely to be {\em list $k$-coloring}:
given at each vertex of a graph a list of $k$ colors chosen from some
larger set, find a coloring of the whole graph in which each vertex
color is chosen from the corresponding list~\cite{JenTof-95}.

\begin{corollary}
We can solve the $3$-coloring and $3$-list coloring problems in time
$O(\Lambda^n)$, the $3$-edge-coloring
problem in time $O(\Lambda^m)$,
and the $3$-SAT problem in time $O(\Lambda^t)$,
\end{corollary}

\begin{corollary}
There is a randomized algorithm which finds the solution to any
solvable $(d,2)$-CSP instance (with $d>3$) in expected time
$O((0.4518d)^n)$.
\end{corollary}

\begin{proof}
Randomly choose a subset of four values for each variable and apply our
algorithm to the resulting $(4,2)$-CSP problem.  Repeat with a new
random choice until finding a solvable $(4,2)$-CSP instance. The random
restriction of a variable has probability $4/d$ of preserving
solvability so the expected number of trials is $(d/4)^n$.
Each trial takes time $O(\Lambda^{(2-\epsilon)n})\approx O(1.8072^n)$.
The total expected time is therefore $O((d/4)^n 1.8072^n)$.
\end{proof}

\section{Vertex Coloring}

\begin{figure}[p]
$$\includegraphics[width=3.25in]{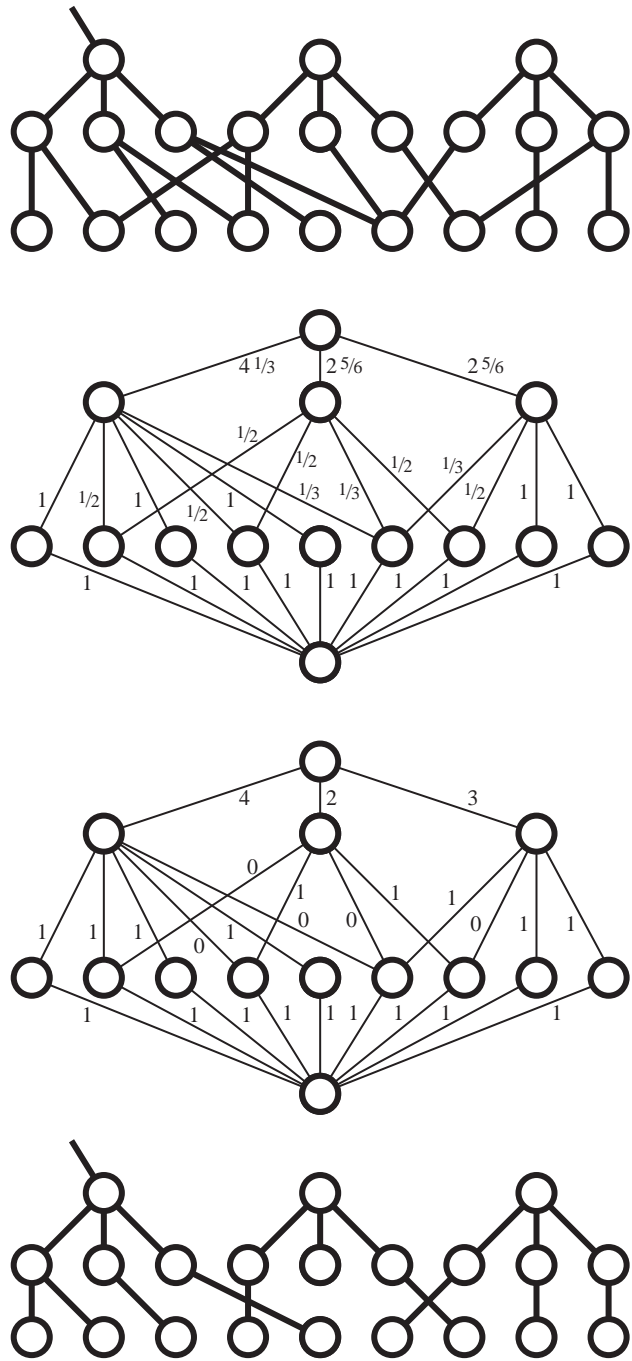}$$
\caption{Use of maximum flow to find a good height-two forest. Top:
forest of $K_{1,3}$ subgraphs and adjacent vertices. The left subgraph
is adjacent to the maximal bushy forest; the other two subgraphs are
not. Top middle: flow graph and fractional flow formed by spreading flow
equally from each of the bottom vertices. The edge capacities are all
one, except for the top three which are respectively 5, 3, and 3.  Bottom
middle: maximum integer flow for the same flow graph.  Bottom: height-two
forest corresponding to the given integer flow.}
\label{fig:flow}
\end{figure}

Simply by translating a $3$-coloring problem into a $(3,2)$-CSP instance, as
described above, we can test $3$-colorability in time $O(\Lambda^n)$.
We reduce this further (as in~\cite{BeiEpp-FOCS-95})
by finding a small set of vertices $S\subset
V(G)$ with a large set $N$ of neighbors, and choosing one of the $3^{|S|}$
colorings for all vertices in $S$.  For each such coloring, we translate
the remaining problem to a $(3,2)$-CSP instance.  The vertices in $S$ are
already colored and need not be included in the $(3,2)$-CSP instance,
and the vertices in $N$ can also be
eliminated, so the overall time is $O(3^{|S|} \Lambda^{|V(G)-S-N|})$.
By choosing $S$ appropriately we can make this quantity smaller than
$O(\Lambda^n)$.

We begin by showing a cycle of degree-three vertices allows us to
reduce a $3$-coloring instance to smaller instances with work factor
$\rec{5,6,7,8}\approx 1.2433$, and that a tree of eight or more such
vertices leads to work factor $\rec{2,5,6}\approx 1.3247$.
Therefore, we can assume that the degree-three vertices form trees of at
most seven vertices, and that the graph contains many higher degree
vertices.

We define a {\em bushy forest} to be an unrooted forest within a given
instance graph, such that each internal node has degree four or more.
Because each tree of $k$ degree-three vertices must have at least $9k/7$
outgoing edges, we can show that
the number of vertices excluded from a maximal bushy forest is at most
$20r/3$, where $r$ is its number of leaves.

All internal nodes of the maximal bushy forest $F$ will be included in
$S$, but we also wish to include some of the remaining graph vertices.
To do this, we form these vertices into trees of height two,
with at most five grandchildren.  Further, in a tree with four or
more grandchildren, at least one node must have degree four or more in
$G$. This forest of height-two trees can be found by the following
flow-based technique (Figure~\ref{fig:flow}): first, find a set of
disjoint $K_{1,3}$ subgraphs in $G\setminus F$, maximal under operations
that remove one such subgraph and form two or more from the remaining
vertices. Assign grandchildren to these height-one trees, from the
remaining vertices nonadjacent to $F$, with fractional
weights spread evenly among the trees each child can be assigned to. This
fractional assignment can be shown to have the bounds we want on the
total assigned weight of grandchildren per tree, but does not form a
disjoint set of trees.  However, we can represent the possible
assignments of grandchildren to trees using a flow graph, and use the
fact that every flow problem has an optimal integer solution, to find a
non-fractional assignment of grandchildren to trees with the same
bounds.  We will add to $S$ certain tree roots or their children,
depending on the shape of each height-two tree. 

\begin{theorem}
We can solve the $3$-coloring problem in time
$O((2^{3/49} 3^{4/49} \Lambda^{24/49})^n)
\approx 1.3289^n$.
\end{theorem}

\begin{proof}
As described, we find a maximal bushy forest
$F$, then cover the remaining vertices by height-two trees.
We choose colors for each internal vertex in $F$,
and for certain vertices in the height-two trees.  Vertices adjacent to
these colored vertices are restricted to two colors, while the remaining
vertices form a $(3,2)$-CSP instance and can be colored using our general
$(3,2)$-CSP algorithm.

Let $p$ denote the number of vertices that are roots in $F$; $q$ denote
the number of non-root internal vertices;
$r$ denote the number of leaves of $F$; $s$ denote the number
of vertices adjacent to leaves of $F$; and $t$ denote the number
of remaining vertices, which must all be degree-three vertices in the
height-two forest.
We show that, if we assign cost $\Lambda$ to each vertex adjacent to a
leaf of $F$, then the cost of coloring each height-two tree, averaged
over its remaining vertices, is at most $(3\Lambda^3)^{1/7}$ per vertex.
Therefore, the total time for the algorithm is at most
$3^p 2^q \Lambda^s (3\Lambda^3)^{t/7}$.

This bound is subject to the constraints
$p,q,r,s,t\ge 0$, $p+q+r+s+t=n$,
$4p+2q\le r$,
$2r\ge s$, and
$20r/3\ge s+t$.
The worst case occurs when $s+t=20r/3$, $s=2r$, $t=14r/3$,
$4p+2q=r$, $p=0$, and $r=2q$, giving the stated bound.
\end{proof}

\section{Edge Coloring}

We now describe an algorithm for finding edge colorings of undirected
graphs, using at most three colors, if such colorings exist.
We can assume without loss of generality that the graph has vertex
degree at most three.
Then $m\le 3n/2$, so by applying our vertex coloring algorithm to the
line graph of $G$ we could achieve time bound
$1.3289^{3n/2}\approx 1.5319^n$.  Just as we improved our vertex coloring
algorithm by performing some reductions in the vertex coloring model
before treating the problem as a $(3,2)$-CSP instance, we improve this
edge coloring bound by performing some reductions in the edge coloring
model before treating the problem as a vertex coloring instance.

\begin{figure}[t]
$$\includegraphics[width=3in]{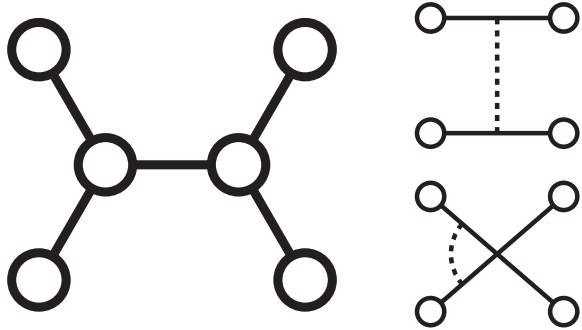}$$
\caption{Replacement of five edges (left) by two constrained edges
(right).}
\label{fig:splice}
\end{figure}

The main idea is to solve a problem intermediate in generality
between $3$-edge-coloring and $3$-vertex-coloring: $3$-edge-coloring with some
added constraints that certain pairs of edges should not be the same
color.

\begin{lemma}\label{lem:splice}
Suppose a constrained $3$-edge-coloring instance contains an unconstrained
edge connecting two degree-three vertices.
Then the instance can be replaced by two smaller instances with three
fewer edges and two fewer vertices each.
\end{lemma}

This reduction operation is depicted in Figure~\ref{fig:splice}. 

We let $m_3$ denote the number of edges with three neighbors in an
unconstrained $3$-edge-coloring instance, and $m_4$ denote the number of
edges with four neighbors.  Edges with fewer neighbors can be removed at
no cost, so we can assume without loss of generality that $m=m_3+m_4$.
By using a maximum matching algorithm, we can find a set of
$m_4/3$ edges such that Lemma~\ref{lem:splice} can be applied
independently to each edge.

\begin{theorem}
We can $3$-edge-color any $3$-edge-colorable graph, in time
$O(2^{n/2})$.
\end{theorem}

\begin{proof}
We apply Lemma~\ref{lem:splice} $m_4/3$ times, resulting in a 
set of $2^{m_4/3}$ constrained
$3$-edge-coloring problems each having only $m_3$ edges.  We then treat
these remaining problems as $3$-vertex-coloring problems on the
corresponding line graphs, augmented by additional edges
representing the constraints added by Lemma~\ref{lem:splice}.
The time for this algorithm is thus at most
$O(1.3289^{m_3} 2^{m_4/3})$.
This is maximized when $m_4=3n/2$ and $m_3=0$.
\end{proof}

\small
\bibliographystyle{abuser}
\let\oldbib\thebibliography
\def\thebibliography#1{\oldbib{#1}\itemsep 0pt}

\bibliography{3color}

\end{document}